# Time-Reversal-Symmetry-Breaking Superconductivity in Epitaxial Bismuth/Nickel Bilayers

Xinxin Gong[1,2], Mehdi Kargarian[3], Alex Stern[1], Di Yue[2], Hexin Zhou[2], Xiaofeng Jin[2], Victor M. Galitski[3], Victor M. Yakovenko[3] and Jing Xia[1]*

[1]Department of Physics and Astronomy, University of California, Irvine, CA, 92697, USA

[2]Department of Physics, Fudan University, Shanghai, 200433, China

[3]Department of Physics, CMTC and JQI, University of Maryland, MD, 20742, USA

(* xia.jing@uci.edu)

Superconductivity that spontaneously breaks time-reversal symmetry (TRS) has been found, so far, only in a handful of 3D crystals with bulk inversion symmetry. Here we report an observation of spontaneous TRS breaking in a 2D superconducting system without inversion symmetry: the epitaxial bilayer films of bismuth and nickel. The evidence comes from the onset of the polar Kerr effect at the superconducting transition in the absence of an external magnetic field, detected by the ultrasensitive loop-less fiber-optic Sagnac interferometer. Because of strong spin-orbit interaction and lack of inversion symmetry in a Bi/Ni bilayer, superconducting pairing cannot be classified as singlet or triplet. We propose a theoretical model where magnetic fluctuations in Ni induce superconducting pairing of the $d_{xy} \pm id_{x^2-y^2}$ orbital symmetry between the electrons in Bi. In this model the order parameter spontaneously breaks the TRS and has a non-zero phase winding number around the Fermi surface, thus making it a rare example of a 2D topological superconductor.

Unconventional superconductors, characterized by spontaneous breaking of additional symmetries other than U(1) gauge symmetry, often have novel properties and reveal new physics. Spontaneous time-reversal symmetry breaking (TRSB) in a superconducting state is one of the most fascinating phenomena. Such superconductors may host unusual particles called Majorana fermions(*1*), much discussed for topological quantum computing(*2*). However, up to now, spontaneous TRSB has been observed only in a few 3D superconductors with bulk inversion symmetry, such as $Sr_2RuO_4$, $UPt_3$, and $URu_2Si_2$(*3-5*). The possibility of spontaneous TRSB has not been investigated experimentally, so far, for a 2D surface superconductivity at interfaces between non-superconducting materials(*6*), which currently attract a lot of interest for possible device applications. Such 2D superconducting systems typically lack inversion symmetry because of the difference between the top and bottom surroundings of the layer. Due to spin-orbit interaction, the spin degeneracy of the electronic states is lifted in such noncentrosymmetric materials, thus making the usual singlet-triplet classification of superconducting pairing inapplicable(*7*). Recent progress in the fabrication of high-quality heterostructures enables artificial realization of such exotic superconducting states. In this paper, we report an observation of spontaneous TRSB in the superconducting state of a 2D epitaxially-grown bilayer made from non-superconducting elements Bi and Ni.

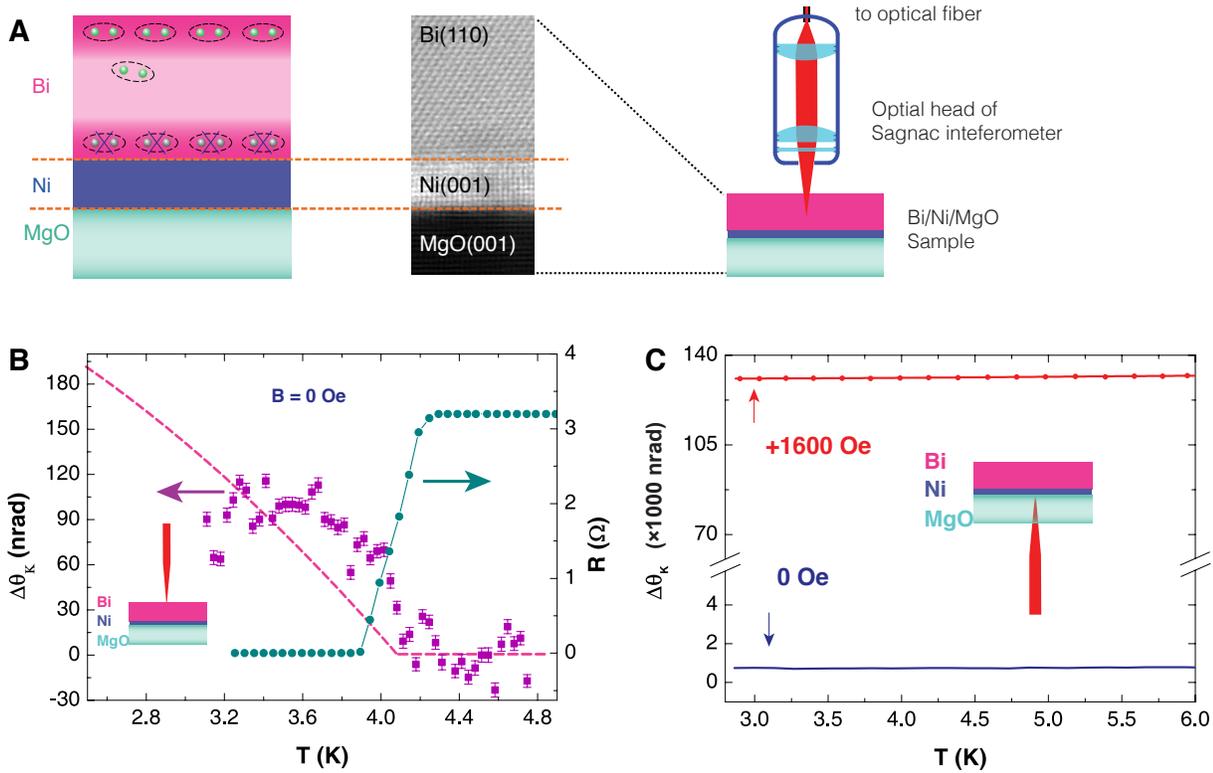

**Fig. 1. Structure and Kerr Signal of a Bi(25 nm)/Ni(2 nm) sample.** **(A)** Left: Side view of the sample structure with TEM image. Right: Schematic diagram of Kerr rotation measurement. **(B)** Kerr angle (red) and resistance (green) measured on the Bi side for zero-field cool-down, showing the onset of $\theta_K$ at $T_c$ = 4.1 K. The red dashed line is a guide to the eye with the form of $1-(T/T_c)^2$. **(C)** The Kerr signal measured on the Ni side, at 1600 Oe applied perpendicular magnetic field (red) or at zero field (blue). In either case, no change of the Kerr signal was observed across $T_c$ = 4.1 K.

The bulk rhombohedral crystalline Bi, which is the heaviest nonradioactive element in the periodic table and a semimetal with an extremely long mean free path, is not superconducting down to 50 mK(*8*). Thus, it was a complete surprise when tunneling experiments(*9, 10*) identified a superconducting transition at $T_c$ = 3.6 K in Bi thin films evaporated on a thin seed layer of ferromagnetic Ni. It is unknown what the symmetry of this superconducting pairing is and, in particular, whether it breaks the TRS. Due to recent progress in epitaxial film technology, it became possible to fabricate high-quality superconducting Bi/Ni films(*11*) suitable for optical measurements of the polar Kerr effect (PKE). The PKE can be utilized for experimental detection of spontaneous TRSB at a superconducting transition, as it has been demonstrated for $Sr_2RuO_4$, $UPt_3$, and $URu_2Si_2$(*3-5*). The PKE measures the optical phase difference between two opposite circular polarizations of light reflected from sample surface, thus giving information about the TRS of the system. For optical frequencies $\omega$, which are much higher than the superconducting gap $\Delta$, the Kerr signal $\theta_K$ is expected(*4*) to be proportional to $(\Delta/\omega)^2$ and is generally less than 1 μrad ($10^{-9}$ radian)(*3, 4*). To measure such a small $\theta_K$, we use a custom-made ultrasensitive loop-less fiber-optic Sagnac interferometer(*12, 13*) (see fig. S1 in Supplementary Materials). The geometry and design of the Sagnac interferometer(*12*) guarantee that all the time-reversal-invariant effects, such as optical birefringence, temperature fluctuations and mechanical vibrations, cancel out, and a non-zero signal $\theta_K$ appears only if the measured sample breaks the TRS. This method(*14, 15*) probes the sample at an optical penetration depth of a few tens of nm and is ideally suited for studying thin films.

The Bi(110)/Ni(001) samples were fabricated using molecular beam epitaxy on MgO (001) substrate (see fig. S2 in Supplementary Materials) and then mounted in an optical cryostat, with the optical probe beam incident normally to the sample surface for the PKE measurements, as shown in Fig. 1A. Four gold wires were attached to the sample surface with silver epoxy for transport measurements. A study of thickness dependence (see fig. S4 in Supplementary Materials) shows that an increase in the thickness of the fcc Ni layer suppresses $T_c$, whereas an increase in Bi thickness restores $T_c$ back to 4.1 K. These observations suggest(*11*) that, while the presence of Ni is necessary for providing the pairing interaction in Bi, superconductivity does not develop at the direct Bi/Ni interface due to the strong ferromagnetism of Ni, but, instead, develops at the opposite surface of Bi facing the vacuum. A further increase of Bi thickness starts to suppress $T_c$(*11*), as expected for going toward the limit of bulk Bi. And there is an optimal Bi layer thickness for maximal $T_c$. In this experiment, we choose a Bi thickness of about 20 nm to minimize the Kerr signal from the underlying Ni layer while maintaining the optimal $T_c$=4.1 K.

Our experimental setup allows us to perform optical measurements on both the Bi and Ni sides of the bilayer, as shown in Figs. 1B and 1C. The main results of the paper are the measurements on the Bi side shown in Fig. 1B. The Bi (25nm)/Ni (2nm) sample was cooled in zero magnetic field (< 0.1 Oe) to the base temperature of the cryostat and then measured upon warming in zero field (ZF). The onset of the ZF Kerr signal is observed at 4.1 K, which coincides with the superconducting transition, as signaled by the initial drop of resistance, thus demonstrating that spontaneous TRSB is due to superconductivity. The Kerr signal $\theta_K$ becomes larger at lower temperatures, reaching 120 nrad at $T$ = 3 K. The dashed line is a guide to the eye of the form of $1-(T/Tc)^2$ (*16*). Each Kerr angle $\theta_K$ data point represents an average of 1000 seconds with the error bar denoting the statistical uncertainty of less than 10 nrad. The systematic source of error/uncertainty is mainly from the long-term drift of the interferometer, which is estimated to be 5 nrad. The Sagnac interferometer operates at 2 μW optical power to prevent any optical heating(*3, 4*) of the sample.

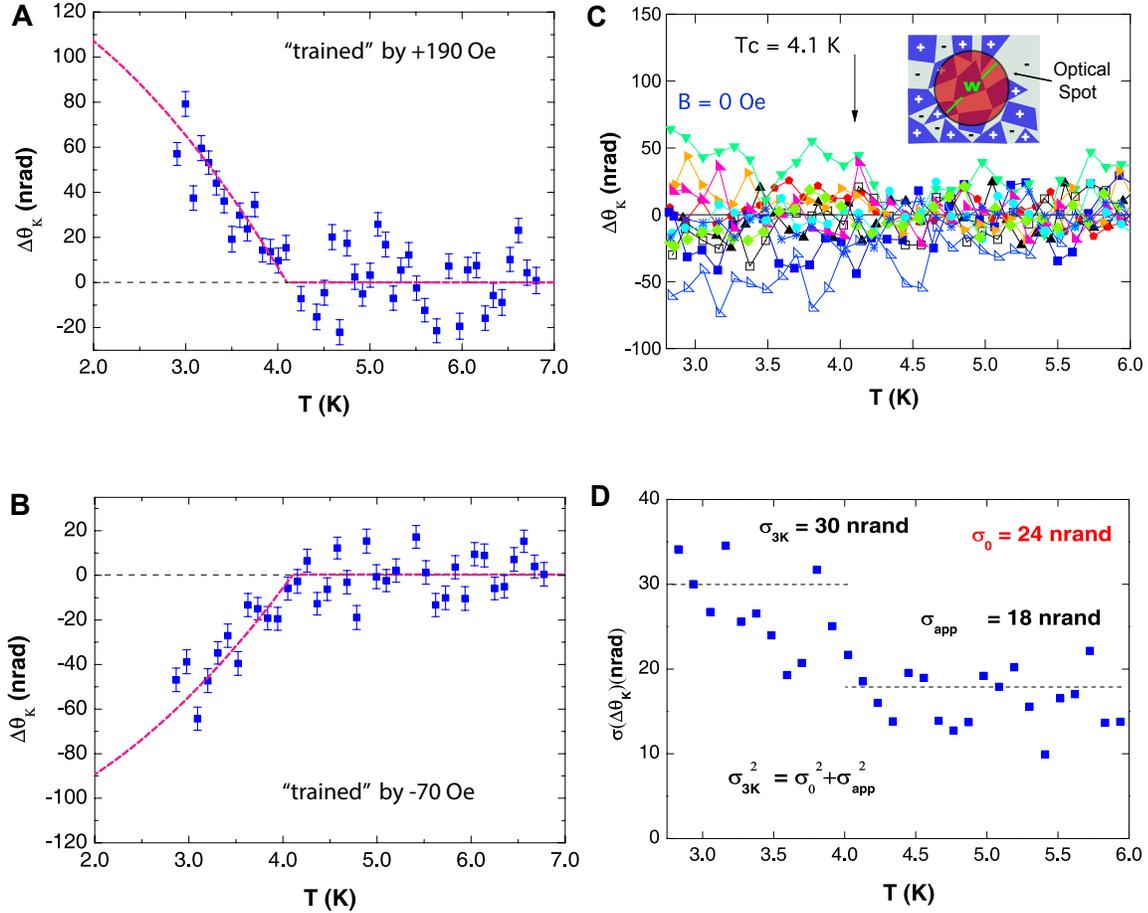

**Fig. 2. Chirality training and domain size estimation of a Bi(20 nm)/Ni(2 nm) sample. (A)** Zero-field warmup data after cooling down in +190 Oe magnetic field. **(B)** Zero-field warmup data after cooling in −70 Oe magnetic field. Red dashed lines are guides to the eye with the form of $1 - (T/T_c)^2$. **(C)** Kerr effect $\theta_K$ measured during zero-field warmup, after cooling down at zero field. The inset is an illustration of random (2D) chiral domains under the optical spot. **(D)** Standard deviation $\sigma(\theta_K)$ between experiments that contains two random contributions: $\sigma_0$ due to chiral domains and $\sigma_{app}$ from the apparatus.

One may ask whether the observed onset of the Kerr signal at 4.1 K on the Bi side of the sample could somehow originate from the ferromagnetic Ni on the other side. We believe it is not the case for the following reasons. First, the 400 K ferromagnetic Curie temperature of Ni is two orders of magnitude greater than the 4.1 K onset temperature of PKE on the Bi side, so it is unlikely that the two effects are related. Second, the magnetic moment of Ni is almost parallel to the interface surface, whereas the PKE is sensitive only to the perpendicular component. Third, the opacity and thickness of the Bi layer greatly suppress an optical pick-up of a signal originating from the Ni side. Nevertheless, to rule out a possibility that a yet-unknown magnetic transition in the underlying Ni layer at 4.1 K generates the Kerr signal on the Bi side, we have directly measured the PKE on the Ni side by flipping the sample as shown in Fig. 1C. Thin Ni films are known to have an almost in-plane magnetic anisotropy, hence the magnetic moment of Ni is almost in the plane, resulting in near-zero polar Kerr signal. This is shown by the blue curve in Fig. 1C, where a temperature-independent $\theta_K$ of 0.75 µrad is observed without any sign of transition at 4.1 K. When a strong magnetic field of 1600 Oe is applied perpendicularly to the interface, the majority of the magnetic moment of Ni is out of the plane. And the corresponding perpendicular component produces a Kerr signal of 130 µrad, as shown by the red curve in Fig. 1C. These indicate a magnetic anisotropy angle of less than 0.33 degree from in-plane direction, which is common for thin Ni films(*10*). During the Kerr measurement on the Bi side, since only a tiny fraction of the light could penetrate into the underlying Ni layer the Kerr signal is less than 5 µrad, or 4% of the Kerr signal from the Ni side. Therefore if the 120 nrad Kerr signal in Fig. 1B were due to an unknown magnetic transition in Ni at 4.1 K, it would necessarily require an increase of the Kerr signal on the Ni side by 3 µrad below 4.1 K, which is non-existent in the blue curve in Fig. 1C to a precision of 0.1 µrad.

Due to spontaneous symmetry breaking, the sign of the Kerr signal observed below $T_c$ may be positive or negative, but it can be controlled by applying a small external "training" magnetic field when cooling through the superconducting transition. Such a training effect has been observed in other TRSB superconductors*(3-5)*, and we also find it in Bi/Ni. As shown in Figs. 2A and 2B, a Bi (20 nm)/Ni (2 nm) sample was cooled down to the lowest temperature in a magnetic field of +190 Oe or -70 Oe perpendicular to the film, and then the Kerr signal was measured on warm-up in zero magnetic field. We observed that the sign of the Kerr signal on warm-up corresponds to the sign of the training field applied during cool-down, thus indicating that the TRSB order parameter couples to the external magnetic field. We have repeated these measurements with training fields up to 230 Oe and found the absolute values of $\theta_K$ at 2.8 K to be identical within the 20 nrad measurement uncertainty. The fact that $|\theta_K|$ is independent of the cooling field magnitude implies that the trapped superconducting flux is not responsible for the observed $\theta_K$, because the number of trapped vortices would scale with the cooling field. The applied training magnetic fields are also too small to alter the small out-of-plane component of Ni magnetic moment.

For a cool-down in zero magnetic field, we expect that chiral domains with positive (+) and negative (-) signs of the Kerr signal should coexist in the sample. If the optical spot is positioned within a single chiral domain, the full Kerr signal would be observed, which we believe is the case for the measurement shown in Fig. 1B performed without a training field. However, more often the measured $\theta_K$ is an average of chiral domains under the focused optical spot of the diameter $w \approx 2$ µm, with a value between $+\theta_0$ and $-\theta_0$, where $\theta_0$ = 60 nrad is the value for a single domain, i.e. after being fully trained. This is, indeed, observed in 11 measurements of the Bi (20 nm)/Ni (2 nm) sample shown in Fig. 2C. The standard deviation $\sigma(\theta_K)$ between these measurements has two contributions: $\sigma_0$ due to random chiral domains and $\sigma_{app}$ from the apparatus uncertainty. Above $T_c$, there is no TRSB, and $\sigma(\theta_K)$ remains constant at $\sigma_{app}$ = 18 nrad. Below $T_c$, the TRSB occurs, and $\sigma(\theta_K)$ increases towards $\sigma_{3K}$ = 30 nrad at $T$ = 3 K as shown in Fig. 2D. Therefore $\sigma_0$ can be estimated to be 24 nrad at $T$ = 3 K. Assuming 2D domains with the average size $d$, we estimate that $d \approx w\, \sigma_0 / \theta_0 \approx 0.8$ µm.

To explain the experimental results, we propose a minimal model involving electronic surface states of Bi interacting with fluctuating magnetic moments of Ni (technical details are given in the Supplementary Materials). There are indications(*17*) that electronic states in the interior of a Bi film are gapped, and the only metallic states are the surface states. So, we only consider the Bi surface exposed to vacuum, because the surface exposed to Ni does not contribute to superconductivity, according to ref. (*11*). The (110) surface of Bi may contain multiple Fermi pockets(*18-22*), but we only consider the largest pocket enclosing the surface Brillouin zone center $\bar{\Gamma}$. Due to the strong Rashba-type spin-orbit coupling, the electron spin and momentum are locked on this pocket, as shown in Fig. 3A. To simplify the presentation, we assume rotational symmetry in the (*x, y*) plane of the surface (anisotropy is discussed at the end) and use the standard Rashba Hamiltonian $v_F(\mathbf{k}\times\boldsymbol{\sigma})\cdot\hat{\mathbf{z}}$, where $\mathbf{k}$ and $\boldsymbol{\sigma}$ are the momentum and spin operators for 2D electrons. The Zeeman term $-\mathbf{M}\cdot\boldsymbol{\sigma}$ due to the in-plane magnetization $\mathbf{M}$ of the Ni can be eliminated by the gauge transformation $\Psi = e^{-i\mathbf{w}\cdot\mathbf{r}}\Psi'$ of the electron fields, which is equivalent to a redefinition of the electron momentum $\mathbf{p} = \mathbf{k} - \mathbf{w}$ and a shift of the Fermi surface center from $\bar{\Gamma}$ to $\bar{\Gamma}'$ by the vector $\mathbf{w} = (\mathbf{M}\times\hat{\mathbf{z}})/v_F$, as shown in Fig. 3A. Then we consider the superconducting pairing of the electrons with the opposite momenta $\mathbf{p}$ and $-\mathbf{p}$ relative to $\bar{\Gamma}'$, as shown in Fig. 3B, which corresponds to a pairing with the total momentum $2\mathbf{w}$ relative to $\bar{\Gamma}$.

Due to the strong spin-orbit interaction and the lack of inversion symmetry (signified by the $\hat{\mathbf{z}}$ direction perpendicular to the surface), the electron states are nondegenerate for each momentum $\mathbf{p}$. Symmetry classification of 2D superconducting pairing in this case(*23*) differs substantially from the conventional, spin-degenerate case. In particular, classification into singlet and triplet pairing is not appropriate(*24*). With the spin quantization axis taken along $\hat{\mathbf{z}}$, the electron state of the momentum $\mathbf{p}$ is the spinor $|\mathbf{p}\rangle = (1, ie^{i\phi_\mathbf{p}})/\sqrt{2}$, where $\phi_\mathbf{p}$ is the azimuthal angle of the vector $\mathbf{p}$ in Fig. 3A. The time-reversed state $|\widetilde{\mathbf{p}}\rangle = \Theta|\mathbf{p}\rangle$ produced by the time-reversal operator $\Theta = i\sigma^y\mathcal{K}$, where $\mathcal{K}$ is the complex conjugation and $i\sigma^y$ is the spinor metric tensor, has the momentum $-\mathbf{p}$, so $|\widetilde{\mathbf{p}}\rangle = \eta_\mathbf{p}|-\mathbf{p}\rangle$, where $\eta_\mathbf{p} = -ie^{-i\phi_\mathbf{p}}$ is a phase factor. Since $\Theta^2 = -1$ for spin 1/2, the phase factor has the property $\eta_\mathbf{p} = -\eta_{-\mathbf{p}}$ (which can also be checked using $\phi_{-\mathbf{p}} = \phi_\mathbf{p} + \pi$) and cannot be eliminated by any gauge transformation. The corresponding second-quantized operators also involve this phase factor: $\tilde{\psi}_\mathbf{p} = \Theta\psi_\mathbf{p}\Theta^{-1} = \eta_\mathbf{p}^*\psi_{-\mathbf{p}}$. Following ref. (*23*), the superconducting condensate $f(\mathbf{p}) = \langle\psi_\mathbf{p}\tilde{\psi}_\mathbf{p}\rangle$ is introduced for the fermion fields that are time-reversal partners of each other. An important observation is that $f(\mathbf{p})$ is an even function of $\mathbf{p}$: $f(\mathbf{p}) = \eta_\mathbf{p}^*\langle\psi_\mathbf{p}\psi_{-\mathbf{p}}\rangle = -\eta_\mathbf{p}^*\langle\psi_{-\mathbf{p}}\psi_\mathbf{p}\rangle = -\eta_\mathbf{p}^*\eta_{-\mathbf{p}}\langle\psi_{-\mathbf{p}}\tilde{\psi}_{-\mathbf{p}}\rangle = \langle\psi_{-\mathbf{p}}\tilde{\psi}_{-\mathbf{p}}\rangle = f(-\mathbf{p})$, where fermion anticommutation was used. Thus, only **even** values of $m$ are permitted in the expansion $f(\mathbf{p}) = \sum_m f_m e^{im\phi_p}$ over circular harmonics for a rotationally-invariant system (because $\phi_{-\mathbf{p}} = \phi_\mathbf{p} + \pi$). The simplest case $m = 0$, studied in ref. (*24*), represents pairing that is rotationally and time-reversal invariant. Importantly, the $p_x \pm ip_y$ pairing with the odd $m = \pm 1$, often discussed for Sr$_2$RuO$_4$, is forbidden in the nondegenerate case.

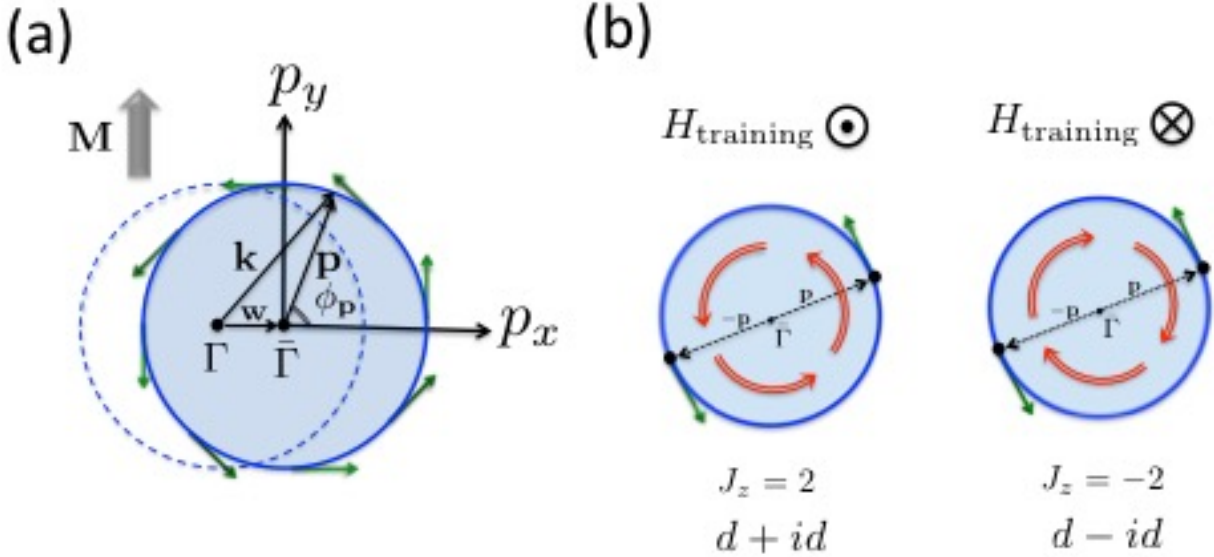

**Fig. 3. Shifted Fermi surface and pairing symmetry of Bi electrons. (A)** The Fermi circle of the electron surface states in Bi. The original dashed-blue Fermi circle centered at $\bar{\Gamma}$ is shifted by the vector $\mathbf{w}$ to the solid-blue circle (shaded area) centered at $\bar{\Gamma}'$ due to the in-plane magnetization $\mathbf{M}$ produced by Ni. The electron momenta $\mathbf{p}$ are measured from $\bar{\Gamma}'$ and characterized by the azimuthal angle $\phi_p$. The green arrows show spin polarization locked to the momentum direction. **(B)** Superconducting pairing of the electrons with opposite spins and opposite momenta $\mathbf{p}$ and $-\mathbf{p}$. The time-reversal-breaking condensate has the total angular momentum $J_z = \pm 2$, corresponding to $d \pm id$ pairing, as indicated by the double red curved arrows. A weak training magnetic field can select one of the two degenerate states.

To explain spontaneous TRSB in Bi/Ni bilayers, we propose that the superconducting pairing in this system takes place with $m = \pm 2$, where the sign is either selected spontaneously or controlled by a small training magnetic field, as shown in Fig. 3B. The pairing states with $m = \pm 2$, which can be also labeled as $d_{xy} \pm id_{x^2-y^2}$, are the time-reversal partners of each other. As shown in ref. *(23)*, these states have two chiral Majorana edge modes propagating around the sample edge either clockwise or counterclockwise. In the conventional spinor basis, the superconducting condensate $F_{\alpha\beta}(\mathbf{p}) = \langle \Psi_{\mathbf{p}\alpha}\Psi_{-\mathbf{p}\beta}\rangle$ is a 2×2 matrix labeled by spin indices $\alpha, \beta =\uparrow, \downarrow$. Although the spin $S_z$ and orbital angular momentum $L_z$ have different values for different components of this matrix, representing a mix of singlet and triplet pairing, the total angular momentum has the same value $J_z = S_z + L_z = m = \pm 2$ for all matrix components. Similar conclusions were obtained in ref. *(25)* using renormalization group for repulsive interaction between electrons. In a less convenient representation employed in ref. *(26)* and ref. *(27)*, the pairing function $\langle\psi_\mathbf{p}\psi_{-\mathbf{p}}\rangle = \eta_\mathbf{p} f(\mathbf{p})$ has the winding number $m - 1$ with the values $1$ and $-3$ (corresponding to p- and f-wave pairing) for the time-reversed pairings with the total angular momentum $J_z = m = \pm 2$.

It was shown in ref. (*28*) that phonons cannot possibly produce the TRSB superconductivity. So, we propose that the superconducting pairing of the surface electron states in Bi layer is mediated by magnetic fluctuations in Ni layer. This scenario is supported by the observation (*11*) that superconductivity exists only for very thin Ni layers and is suppressed with the increase of Ni thickness, which would reduce magnetic fluctuations. We particularly focus on the out-of-plane fluctuations of the Ni magnetic moments (the effect of in-plane fluctuations is discussed in ref. (*29*)). The effective interaction in the Cooper channel between electrons with the momenta $\mathbf{p}$ and $\mathbf{p}'$ is given by the properly symmetrized product of the magnon propagator $D(\mathbf{p} - \mathbf{p}')$ and the spin form-factor $\Lambda^s_{\mathbf{p},\mathbf{p}'}$ (*30*). The dominant contribution to the $m = \pm 2$ pairing channel comes from the product of the first circular harmonics of $D$ and $\Lambda^s$, but is somewhat reduced by the product of the second harmonic of $D$ and zeroth harmonic of $\Lambda^s$.

Finally, we discuss the effects of crystal anisotropy. The (110) surface of Bi has a quite low symmetry group **D**$_1$ with a mirror plane perpendicular to the surface (*18*). Consequently, there are only two symmetry representations $A_1$ and $A_2$ for the superconducting pairing (*23*), which are symmetric and antisymmetric with respect to reflection. The antisymmetric representation $A_2$ has nodes and is even in $\mathbf{p}$, so its wave function is $p_x p_y$ of the $d_{xy}$ type. The symmetric representation $A_1$ can be of the $s$ or $d_{x^2-y^2}$ type, which belong to the same representation. Importantly, because two different representations $A_1$ and $A_2$ are available, it is possible to make two degenerate time-reversal-breaking combinations $d_{xy} \pm i d_{x^2-y^2}$ out of them. However, because the $A_1$ and $A_2$ representations are not degenerate, their transition temperatures are generally different. So, we come to conclusion that there must be two superconducting transitions in Bi/Ni bilayers, as in UPt$_3$ (*4*). We expect that the nodal superconducting state of the $d_{xy}$ type would develop at a higher transition temperature, and the $d_{x^2-y^2}$ state would develop at a lower temperature, spontaneously breaking the TRS by the $d_{xy} \pm i d_{x^2-y^2}$ combination and eliminating the nodes. If the crystal anisotropy is weak, the difference between transition temperatures may be small, which may be the reason why the two superconducting transitions have not been resolved so far, but it is an important direction for future experimental studies.

In conclusion, we report observation of the polar Kerr effect in Bi/Ni epitaxial bilayers, which indicates spontaneous TRSB in the superconducting state. A combination of strong spin-orbit coupling and noncentrosymmetry makes this 2D interfacial system distinctly different from the bulk TRSB superconductors (*3-5*). To describe the TRSB state, we propose a model where magnetic fluctuations in Ni induce superconducting pairing in Bi with the superconducting phase winding number around the Fermi surface equal $\pm 2$. Thus, Bi/Ni is likely a rare example of a 2D topological superconductor with two chiral edge states moving in the same direction (*23*).

**Materials and Methods**

**Setup of the Sagnac interferometer**

A schematic of the Sagnac interferometer used in this experiment is shown in fig. S1. In a fiber-optic Sagnac interferometer (*13*), light from a 1550 nm broadband source passes through a polarization maintaining (PM) circulator and is polarized. A half-wave plate is then used to rotate the polarization axis to 45 degrees with respect to the axis of a phase shift modulator. This modulator adds a time-varying phase shift to the light. However, the amplitude of the phase shift and the optical path length are different for the in-plane and out-of-plane polarizations of light. After passing through the modulator, the in-plane and out-of-plane polarizations are no longer coherent with each other, and each polarization beam has a time varying phase shift of different amplitude. The light is then routed via a PM fiber to optics mounted on a piezo stage capable of sub-micron step size in both the x- and y-directions. After exiting the PM fiber, the light is passed through a quarter-wave plate which converts the orthogonal, linear polarizations into left- and right-circular polarizations. Finally, it is focused onto the sample surface. The sample rests inside an optical cryostat with a base temperature of 2.5 K, which contains an optical viewport on the top allowing the light to pass from the focusing element to the sample surface. Upon reflection from the sample, there is a phase shift between the two polarizations equal to the twice the Kerr signal. After passing through the quarter-wave plate the polarization axis of the two beams had been swapped. On the return trip each beam travels through the system with a polarization orthogonal to that of its outgoing trip. The net result is that each beam travels along the exact same optical path but in opposite direction over the round trip. Once each beam passes through the phase shift modulator again, the two beams are once again coherent but have phases differences due to the Kerr effect ($2\theta_K$) as well as that caused by the phase shift modulation [$\varphi_m(\sin(\omega_m(t+\tau)) - \sin(\omega_m t))$, where $\omega_m$ is the modulation frequency and $\varphi_m$ is the difference in amplitudes between the in-plane and out-of-plane modulation]. The two beams interfere, resulting in an elliptically polarized beam, the in-plane component of which is routed to a photo detector. If $\omega_m$ is chosen such that $\tau\omega_m = \pi$, lock-in detection can be used to determine $\theta_K$ by comparing the detected signal at $\omega_m$ and $2\omega_m$. The DC power received by the photo detector is a measure of the reflectivity of the sample, and the ratio between the signal at $2\omega_m$ and the DC signal is a measure of optical anisotropy as described in reference (*13*).

**Procedure of epitaxially growing samples**

All the samples were epitaxially grown in ultrahigh vacuum (UHV) with a base pressure $6\times10^{-8}$ Pa. We first chemically cleaned the MgO (001) substrates before putting them into the UHV chamber. After transfer, we annealed them to 750 K and stay at this temperature for 70 min. Figure S2A shows the reflection high-energy electron diffraction (RHEED) patterns for the MgO substrate. The Ni layer was deposited at 300 K while the Bi layer at 110 K. Figure S2B shows the RHEED patterns of the topmost Bi layer taken at room temperature. The sharp lines indicate the sample is single crystalline and the quality is very high, which can also be verified by the cross section scanning transmission electron microscopy (TEM) image in Fig. 1A.

**Acknowledgements:** V.M.Y. thanks K. Samokhin for a discussion. **Funding:** We thank the generous support from NSF grant DMR-1350122. X.G., D.Y., H.Z. and X.J. acknowledge support from NSF of China (grant No. 11434003 and No. 11421404) and National Basic Research Program of China (grant No.2015CB921402). M.K. and V.G. acknowledge support from DOE-BES(DESC0001911) and Simons Foundation.
**Author contributions:** X.G. took the responsibility of conducting the measurements and analyzing the data. A.S. assisted the experiment. D.Y. and H.Z. prepared the sample. X.J. developed the Sagnac interferometer. X.J. and J.X. designed the project. M.K. and V.M.Y. conducted the theoretical calculation. J.X., V.M.Y. and V.M.G. wrote the paper. All of the authors contributed to the result analysis and the preparation of the manuscript.

**Supplementary Materials:**

fig. S1. Schematic of Sagnac interferometer.
fig. S2. RHEED patterns of the substrate and the sample.
fig. S3. Supplementary data of the Bi(25 nm)/Ni(2 nm) sample.
fig. S4. Dependence of the critical temperature $T_c$ on Bi and Ni layer thicknesses.
fig. S5. Kerr signal of the Bi(40 nm)/Ni(2 nm) sample.
fig. S6. Optical signal from the MgO substrate
Theory of Superconducting Pairing in Epitaxial Bismuth/Nickel Bilayers
References(*31-41*)

**Supplimental information in pdf can be found in the following URL:**

http://www.physics.uci.edu/~xia/X-lab/Publications_files/Supplementary_Information.pdf